# Baryonic Dark Halos: Machos and Cold Gas ?

Ortwin Gerhard[1] and Joseph Silk[2]

[1] *Astronomisches Institut, Universität Basel, Venusstrasse 7, CH-4102 Binningen, Switzerland*
[2] *Departments of Astronomy and Physics, and Center for Particle Astrophysics, University of California, Berkeley, CA 94720, USA*

**ABSTRACT**
We consider the possibility that the dark matter in the halos of galaxies may be in the form of clusters of Machos within which are embedded cold, dense gas clouds. Microlensing experiments have found evidence that the Galactic halo contains up to half of its mass in the form of low-mass Machos. A number of observational and dynamical arguments point to the existence of hitherto unobserved cold gas around galaxies. We show that the cold gas can be stabilized by Macho clusters. Within the framework of a simple two-component model, we derive constraints on the Macho clusters and on the halo cold gas content. Typical cluster masses are $\sim 10\,\mathrm{M}_\odot$, typical Macho masses are $\sim 0.01\,\mathrm{M}_\odot$, and the gas content could be up to of order 50 %. Various predictions are given for testing the hypothesis that such objects could constitute most of the mass in the dark halos of galaxies. If halos are indeed baryonic and contain significant amounts of cold gas, they are likely to play a much more active role in galaxy formation and evolution than is commonly supposed.

**Key words:**

## 1 INTRODUCTION

That dark halos may be mostly baryonic in composition is a respectable proposition, in view of the indications from primordial nucleosynthesis that there are considerably more baryons to be accounted for than are seen in luminous form (Copi *et al.* 1995; Krauss & Kernan 1995). The growing evidence that a substantial fraction of galaxy cluster mass is in the form of baryons (Briel, Henry & Böhringer 1992, Buote & Canizares 1995) further supports the idea that baryons contribute significantly to dark matter. Indeed, gravitational microlensing experiments have detected dark matter in the form of Machos (Alcock *et al.* 1993, Aubourg *et al.* 1993). However, the implied mass in Machos falls short by a factor of about 5 (Alcock *et al.* 1995a) of that required by the galactic rotation curve in a "standard" dark halo, although uncertainties in the galactic rotation curve and halo mass distribution can considerably weaken this constraint (Alcock *et al.* 1995b).

Since the only type of dark matter detected to date is baryonic, it is natural to ask whether the entire dark halo could be baryonic. Cold gas is the most logical form for the additional dark baryons, given that the microlensing experiments exclude a large mass range that includes both giant planet and brown dwarf masses. Arguments against baryonic dark matter in the form of diffuse gas and low mass stars have been summarized by Hegyi and Olive (1986). However these constraints can be evaded if the gas is sufficiently dense and cold. Pfenniger *et al.* (1994a,b) have even argued that all of the rotation curve might be accounted for by unseen cold gas hidden in an extended galactic disk. However, these authors did not address issues of global stability of such disks, and there are difficulties with the local surface density of the galactic disk (summarized by Kuijken 1995), the flaring of HI disks (van der Kruit & Shostak 1984, Olling & van Gorkom 1995), gravitational instabilities (Elmegreen 1995), and FIRAS limits on emission from very cold dust (Reach *et al.* 1995). Searches for mm-absorption by cold molecular gas in the outer disk have found some cold clouds, but have not seen large amounts of hitherto undetected cold gas (Lequeux, Allen & Guilloteau 1993; Wilson & Mauersberger 1994; Digel, de Geus & Thaddeus 1994).

Here we consider a model in which baryonic dark matter is distributed in a three-dimensional halo. We imagine that early during galaxy formation cold gas clouds collapse and fragment to form clusters of small Jupiter-like objects. While star formation at the present epoch results in formation of stars with characteristic mass $\sim 0.3\,\mathrm{M}_\odot$ (spanning the range $0.08-100\,\mathrm{M}_\odot$), it is an intriguing possibility to imagine that early star formation (in the Galactic halo) may have resulted in the formation of very low mass objects. If so, a natural scale is indicated by theoretical discussions of opacity-limited fragmentation (Rees 1976, Silk 1977). These arguments provide a relatively robust and nearly metallicity-independent lower bound on the minimum fragment mass of $\sim 10^{-3}\,\mathrm{M}_\odot$ in cold molecular gas, although the highly uncertain physics of fragment coalescence and accretion could raise this estimate by an order of magnitude (Silk & Takahashi 1979).

One could plausibly imagine that such a characteristic mass, which differs dramatically from that attained in present-day star formation, could be attained in an environment where the physical conditions were very different from those in nearby molecular clouds. The most obvious difference maybe that of magnetic field strength: magnetic fields play a critical role in helping protostellar fragments transfer angular momentum to their surroundings. A primordial cloud with negligible magnetic flux may fragment into predominantly very low mass blobs, as originally argued by Mestel (1965). The ratio of fragment mass to initial disk mass $M_d$ for a cold, centrifugally–supported collapsing cloud is $v_s^4 M_d^2 \mu^{-2} h^{-4}$, where $v_s$ is the sound velocity, $h$ is the specific angular momentum, and $\mu$ is the surface density, taken to be constant in the 1-dimensional collapse to a cold, gravitationally unstable disk. Unless $h$ can be reduced by several orders of magnitude, rotational support provides a formidable barrier to assembling solar mass stars.

This argument suggests that very low mass objects form in a primordial, collapsing cloud. Of course, other processes such as density waves may help transfer angular momentum in protostellar disks (Larson 1990) and nonlinear interactions, such as coagulation, coalescence, and accretion (Silk 1978), will help drive up the fragment masses. Some massive stars may form, to provide a seeding of heavy elements. Nevertheless, the inhibition of magenetic flux-regulated angular momentum transfer, currently thought to be a crucial element in determining the initial mass function (IMF), argues that the characteristic stellar mass associated with pregalactic star formation is likely to be displaced towards the minimum value of suggested by opacity-limited fragmentation, to $\sim 0.001 - 0.01 \, \mathrm{M}_\odot$.

Such low mass objects provide the most plausible form of halo baryonic dark matter. Deep imaging of high Galactic latitude fields rules out M dwarfs as a substantial contribution to halo mass (Hu *et al.* 1994, Bahcall *et al.* 1994), and so only an upturn in the IMF below $\sim 0.1 \, \mathrm{M}_\odot$ could result in a significant mass contribution to the halo. Thus if a substantial component of halo dark matter is baryonic, one is compelled observationally by the Macho experiment to take very seriously the hypothesis that the halo IMF, formed in the pregalactic era, is very different from the present epoch IMF, and has a characteristic mass well below that of extreme M dwarfs.

Once a sufficient fraction of the cloud has been converted into a cluster of Jupiters (now observed as Machos), the residual gas can be stabilized as we show below. Significant amounts of cold gas could thus reside undetected in the outer halo, providing a gas reservoir for chemical evolution and disk build-up over the age of the galaxy. It also seems likely that Machos would naturally form in clusters, rather than in isolation. Clustering would affect the inferred Macho content of the halo only if the clusters are very massive (Moore and Silk 1995, Alcock *et al.* 1995b).

In this article, we review evidence for some of the halo dark matter being gaseous (Section 2), then consider an idealized model for the stabilization of cold gas by a surrounding collisionless cluster of Machos (Section 3), develop a simple model for two-component halo clouds and their distribution (Sections 4, 5), and finally provide various predictions for testing the hypothesis that halo dark matter may be in this form (Section 6).

## 2   THE CASE FOR BARYONIC DARK MATTER IN GALAXY HALOS

### 2.1   General arguments

#### 2.1.1   *Cosmic nucleosynthesis*

Standard cosmic nucleosynthesis calculations together with the best current observational limits on the abundances of $^2D$, $^3He$, $^4He$, and $^7Li$ constrain the cosmic baryon density (Copi *et al.* 1995; Krauss & Kernan 1995) in the range:

$$0.01 < \Omega_b h^2 < 0.02. \tag{1}$$

Since the Hubble constant, $H_0 = 100h \, \mathrm{km \, s^{-1} \, Mpc^{-1}}$, is believed to be bounded by $0.5 < h < 0.8$, between 1% and 8% of the closure density $\Omega_{\mathrm{crit}}$ must be in the form of baryons. On the other hand, the stellar and gaseous mass of visible galaxies is only $\lesssim 0.5\%$ of $\Omega_{\mathrm{crit}}$. Thus most of the baryonic mass in the universe remains to be accounted for. Coincidentally, the amount of mass estimated for galaxies, including their massive halos, is approximately what would be needed in order to satisfy the above constraint from nucleosynthesis.

#### 2.1.2   *Rotation curves*

It is well known that the total mass of the luminous stars and of the atomic and molecular gas in spiral galaxies fails to explain the amplitude of the observed rotation velocities in the outer parts of spiral galaxies. There are many carefully studied examples of this (see, e.g, Sancisi & van Albada 1987 for a review). However, it is less well known — although already noted by Bosma (1981) and reiterated by Sancisi (1988) — that the shape of the circular velocity curve deduced from the HI gas alone, scaled by a factor of 2-3, comes suspiciously close to the actually observed rotation curve in the outskirts of many of these galaxies. A striking example is the dwarf spiral galaxy DDO 154 (Carignan 1988). This can also be seen in a surprising number of cases found in a recent study of carefully chosen spiral galaxies with especially regular HI velocity fields (Begeman *et al.* 1991). Indeed, the spiral galaxy NGC5907 shows evidence for a faint luminous halo whose surface brightness profile may be consistent with the dark mass distribution inferred from the rotation curve (Sackett *et al.* 1994a).

#### 2.1.3   *Tully-Fisher relation*

The total luminosities of spiral galaxies correlate rather well with their asymptotic circular velocities (Tully & Fisher 1977). The scatter is, in fact, so small that non-circular motions of $\lesssim 10\%$ can be ruled out from statistical investigations of projection effects (Franx & de Zeeuw 1992). Although there have been attempts to explain this very tight correlation of two apparently unrelated physical quantities in terms of the gravitational pull of the assembling disk on the mass distribution of the pre-existing massive halo (Blumenthal *et al.* 1986), these explanations have never been fully convincing. A conceptually simpler explanation would clearly be one in which one and the same (baryonic) component were responsible both for the rotation velocity and (through some star formation stability criterion) also for the observed luminosity.

Through the Tully-Fisher relation, the correlation between maximum rotation velocity and rotation curve slope found in Casertano & van Gorkom (1991) translates into one between slope and total luminosity. This latter correlation should not occur if the luminous and dark matter are well-mixed, unless the angular momentum parameter $\lambda$ of the galaxy correlates with mass, or unless higher-mass systems have lost less of their baryonic mass than lower mass systems in galactic winds. There is little theoretical support for either possibility. An alternative explanation might be that the fraction of baryons that have become luminous is larger in higher luminosity systems.

## 2.2 Cold gas as dark matter

### 2.2.1 *HI velocity fields*

The HI distributions in the outer parts of spiral disks are often lopsided, yet in some cases the gas seems to follow an almost exactly circular rotation velocity field. The latter observation rules out the possibility that the gravitational potential is lopsided, and also that the gas has arrived recently, because the differential precession time-scales are a few $10^9$ yr. The lopsidedness should damp out on this time-scale simultaneously with the non-circular motions. There remains the possibility that the observed gas is not a faithful tracer of the total mass in the disk – i.e., there is unseen baryonic matter hidden in the outer disk. This might be ionized (Maloney 1993), or molecular (Lequeux *et al.* 1993). Two impressive examples are the Sc galaxy NGC 3198 (Begeman 1989) and the edge-on Sb galaxy NGC 891 (Sancisi & Allen 1979, Rupen 1991). In NGC 3198, the HI density contours at low intensities become more symmetric again. This indicates that the asymmetry cannot be caused by the ambient UV radiation field, since this would limit the HI disk at a fixed column density.

### 2.2.2 *Counter-rotating disks*

A somewhat different situation arises with the Virgo S0 galaxy NGC 4550, recently shown to contain two approximately equally massive counterrotating disks (Rubin *et al.* 1992, Rix *et al.* 1992). The formation of a cold second component, while keeping the first disk component as cold as is observed, is difficult to understand unless it occurs adiabatically. This cannot be simply accomplished in the context of the standard cold dark matter cosmology, where the dark matter accreted with a second galaxy-sized lump of baryonic mass tends to overheat the preexisting disk (Toth & Ostriker 1992). On the other hand, if gas is released only slowly from a baryonic halo in a model like that described in Section 5.6, an adiabatic build-up of the second disk is easier to imagine; e.g., if the specific angular momentum of the halo reverses at large radii.

### 2.2.3 *Gas consumption rates in star-forming galaxies*

The present rate of star formation in some spiral galaxies suggests that baryonic infall at a rate of up to several solar masses per year may be required to avoid gas exhaustion within a small fraction of a Hubble time, and therefore the inference of a special present epoch (Sandage 1986, Kennicutt *et al.* 1994). Clearly, a reservoir of diffuse dark baryonic matter with an order of magnitude more mass than seen in stars provides an ideal means of extending disk lifetimes, if a few percent of the dark matter is released over a Hubble time.

Arguments have also been advanced (Braine & Combes 1993) that merging or interacting galaxies contain a higher mass of cold gas per unit blue luminosity than do unperturbed galaxies. While this result is uncertain due to the poorly known conversion of CO emissivity to molecular gas mass, it may also point to extra unseen gas around normal galaxies.

## 3 STABILITY OF GAS SPHERES IN DARK MATTER CLUSTERS

Isolated isothermal and polytropic gas spheres with $\gamma < 4/3$ are unstable to gravitational collapse and would fragment into star-like objects. The physical cause of this instability is that at sufficiently large radii, the internal pressure of the sphere has fallen sufficiently that it can no longer resist the combined effect of an external pressure perturbation and the sphere's self-gravity. However, similar gas clouds can be stabilized by embedding them in a background of collisionless dark matter which reduces the self-gravity of the gas. This was shown for a homogeneous background matter density by Umemura & Ikeuchi (1986). Their result suggests that once a sufficient fraction of a gas cloud has fragmented into Jupiter-like objects, this quasi-collisionless component might then stabilize the remaining gas in the cloud by reducing its self-gravity. Thus in this section we investigate the stability of isotropic and polytropic gas spheres in the background potential of a collisionless cluster mass distribution. The main objective of this analysis is to estimate the mass ratio required for stabilization.

### 3.1 Equilibrium configurations of polytropes bounded by collisionless dark matter

A simple model for a gas cloud gravitationally confined within a Macho cluster is described by the following equations for gravitational and hydrostatic equilibrium:

$$\frac{1}{r^2}\frac{d}{dr}\left(r^2\frac{d\psi}{dr}\right) = -4\pi G \left[\rho_g(r) + \rho_d(r)\right], \tag{2}$$

$$\frac{1}{\rho_g}\frac{dp}{dr} = \frac{d\psi}{dr}, \tag{3}$$

$$p = kT\rho_g/m \equiv \kappa \rho_g^{1+\frac{1}{n}}, \tag{4}$$

where $\rho_g$ and $\rho_d$ are the density distributions of gas and dark matter, presumed to be spherically symmetric, $\psi$ is the (positive) joint gravitational potential, $p$ and $T$ are the gas pressure and temperature, and $n$ is the polytropic index of the gas. As a simple model for the background stellar density distribution we take the density distribution

$$\rho_d(r) = \rho_{d0}\frac{r_c^2 r_t^2}{(r^2 + r_c^2)(r^2 + r_t^2)}, \tag{5}$$

which has constant density well inside the density profile core radius $r_c$ and eventually steepens to $\rho_d(r) \propto r^{-4}$ far outside the second characteristic radius $r_t$.

We denote the central gas density and temperature by $\rho_{g0}$ and $T_0$, and define the usual dimensionless variables $\xi$, $\theta$,

assuming that the polytropic index $n > -1$:

$$r \equiv r_1 \xi = \left(\frac{n+1}{4\pi G}\kappa\rho_{g0}^{\frac{1}{n}-1}\right)^{\frac{1}{2}} \xi, \tag{6}$$

$$\rho_g \equiv \rho_{g0}\theta^n. \tag{7}$$

With equation (4), the radial length scale for the gas can be written

$$r_1 = \left(\frac{n+1}{4\pi G\rho_{g0}}\right)^{\frac{1}{2}} \left(\frac{kT_0}{m}\right)^{\frac{1}{2}}. \tag{8}$$

We now define an equivalent central temperature $T_{d0}$ for the collisionless component by the similar relation

$$r_c \equiv \left(\frac{n+1}{4\pi G\rho_{d0}}\right)^{\frac{1}{2}} \left(\frac{kT_{d0}}{m}\right)^{\frac{1}{2}}. \tag{9}$$

Then we may write

$$\frac{r_1^2}{r_c^2} = \frac{\rho_{d0}}{\rho_{g0}}\frac{T_0}{T_{d0}} \equiv \beta\tau^2, \tag{10}$$

where

$$\beta \equiv \frac{\rho_{d0}}{\rho_{g0}}, \qquad \tau^2 \equiv \frac{T_0}{T_{d0}}. \tag{11}$$

The gravitational equilibrium equation reduces to

$$\frac{1}{\xi^2}\frac{d}{d\xi}\left(\xi^2\frac{d\theta}{d\xi}\right) = -\theta^n - \beta\left(1+\beta\tau^2\xi^2\right)^{-1}\left(1+\beta\tau^2\frac{r_c^2}{r_t^2}\right)^{-1}. \tag{12}$$

This is solved subject to the boundary conditions

$$\theta(\xi=0)=1, \quad \theta'(\xi=0)=0. \tag{13}$$

For a given polytropic index $n$, the solutions depend on the ratio of central densities $\beta$, the ratio of central temperatures $\tau^2$, and the structural parameter $r_c/r_t$ of the collisionless component.

Figure 1 shows a series of equilibrium solutions for polytropic gas of varying central density in such a two-component model. The polytropic index is $n=7$, the temperature ratio $\tau^2=0.8$, and $r_c/r_t=0.5$. The radii and masses of these polytropic spheres are all finite; this is caused by the extra gravitational field of the background component. Isolated $n=7$-polytropes have infinite extent and mass. While finite polytropic solutions can be found in many cases, especially when the gas component is more centrally concentrated than the collisionless cluster, it is clear that infinite solutions also exist, e.g., for sufficiently small $\beta$. For some values of the parameters, the solutions change from finite to infinite with varying central density.

To investigate the stability of these embedded gas spheres, we study series of equilibrium configurations of varying central gas density for a fixed equation of state and dark matter mass. This implies from equation (4) that the central gas temperature varies $\propto \rho_{g0}^{1/n}$ and hence that $\tau^2 \propto \rho_{g0}^{1/n}$ along the sequence. We find the total gas mass $M_g(\rho_{g0})$ and the turning points at which $dM_g/d\rho_{g0}=0$. At such a turning point, the second variation of the energy vanishes, implying the change of stability of one of the eigenmodes of the gas sphere (e.g., Shapiro & Teukolsky 1983).

The gas mass within radius $r$ is given by

$$M_g(r) = -\frac{r^2}{G}\frac{d\psi}{dr} - M_d(r)$$
$$= -4\pi\rho_{g0}r_1^3\xi^2\frac{d\theta}{d\xi} - M_d(r). \tag{14}$$

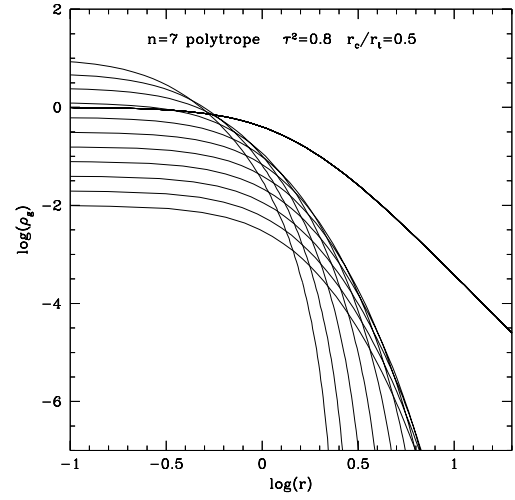

**Figure 1.** Equilibrium polytropes ($n=7$) in the combined gravitational potential of the gas and a collisionless cluster of Machos. The two components have central temperature ratio $\tau^2=0.8$; the gas component is then more centrally concentrated. The typical gas mass in these particular models is 15 – 20% of the Macho cluster mass. These polytropic configurations have finite radius and mass, contrary to their isolated counterparts.

The mass within $r$ of the collisionless component of equation (5) is

$$M_d(r) = \frac{4\pi\rho_{d0}r_c^3 r_t^3}{r_c(r_t^2-r_c^2)}\left[\tan^{-1}\left(\frac{r}{r_t}\right) - \frac{r_c}{r_t}\tan^{-1}\left(\frac{r}{r_c}\right)\right], \tag{15}$$

and the total mass is

$$M_d(\infty) = \frac{2\pi^2\rho_{d0}r_c^3 r_t^2}{r_c(r_t+r_c)}. \tag{16}$$

Using equations (6) and (10)-(11) we may rewrite $M_g(r)$ in terms of dimensionless parameters:

$$\frac{M_g(r)}{M_d(\infty)} = -\frac{2}{\pi}\left[\beta^{\frac{1}{2}}\tau^3\left(1+\frac{r_c}{r_t}\right)\left(\frac{r_c}{r_t}\right)\xi^2\frac{d\theta}{d\xi} + \left(1-\frac{r_c}{r_t}\right)^{-1}\right.$$
$$\left.\times\left\{\tan^{-1}\left(\frac{r_c}{r_t}\beta^{\frac{1}{2}}\tau\,\xi\right) - \frac{r_c}{r_t}\tan^{-1}\left(\beta^{\frac{1}{2}}\tau\,\xi\right)\right\}\right]. \tag{17}$$

Figure 2 shows the $M(\rho_{g0})$-curves for three series of polytropic equilibria of varying temperature ratio $\tau^2$. The polytropic index $n=7$ and the structure of the collisionless cluster have been held fixed; the ratio of central densities varies along the curves. The portions of the curves to the right of their respective turning points are unstable; they correspond to polytropic spheres that, although still finite, are sufficiently self-gravitating. The solutions at the bottom left of the figure correspond to gas spheres of very low density which are completely non-selfgravitating and thus stable. Since there is only one turning point in each curve between these limiting cases, this must therefore correspond to the change of stability of the fundamental mode. Hence the entire sequence of equilibria to the left of the turning point is stable.

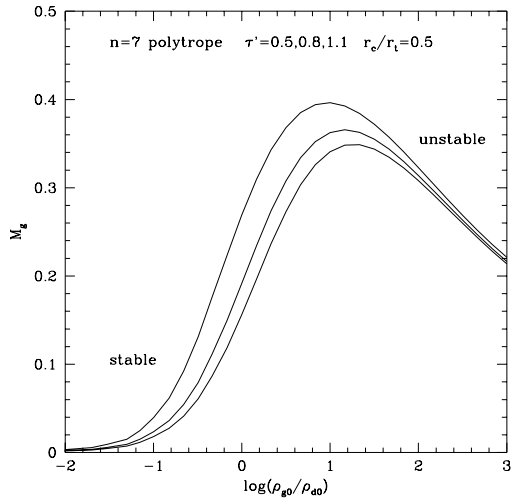

**Figure 2.** Mass-central density relations for three sequences of $n = 7$ polytropes in the combined gravitational potential of the gas and a collisionless cluster of Machos. The central temperature ratio is $\tau^2 = \tau'^2$ at $\rho_{g0}/\rho_{d0} = 1$ where $\tau' = (0.5, 0.8, 1.1)$, and scales $\propto \rho_{g0}^{1/n}$ along each sequence. Polytropes to the left of the respective turning point are stable, those to the right unstable.

| $n \setminus \tau'$ | 0.5 | 0.7 | 0.9 | 1.1 | 1.3 |
|---|---|---|---|---|---|
| 5 | 0.037 | 0.10 | 0.23 | 0.43 | 0.75 |
| 7 | 0.021 | 0.059 | 0.12 | 0.25 | |
| 10 | 0.012 | 0.033 | 0.074 | 0.14 | |

**Table 1.** Maximum mass fraction of polytropic gas embedded in a cluster of collisionless Machos, as function of polytropic index $n$ and temperature parameter $\tau'$ - see Fig. 2. When no entry is given, some members of the corresponding sequence of equilibrium polytropes are infinite.

We have constructed similar sequences of equilibria for a range of polytropic indices and temperature ratios, keeping $r_c/r_t = 0.5$ fixed. For each sequence we have determined the turning point in the $M(\rho_{g0})$ relation and the corresponding ratio of the masses in gas and collisionless matter, $M_g/M_d$. This represents the maximum mass fraction of gas that can be stabilized by the Macho cluster for the corresponding parameters. Table 1 lists these maximum mass ratios. It is seen that $\sim 10 - 50\%$ of the Macho cluster mass can be embedded as a stable sphere of polytropic gas. We may expect these numbers to be increased in more realistic cluster models that include flattening and rotation. We now turn to the global properties of a halo containing many such Macho clusters.

## 4 DISTRIBUTION OF HALO DARK MATTER

### 4.1 Global stability

Cold massive disks are unstable to radial and non-radial instabilities. Prevention of radial instabilities requires a minimum radial velocity dispersion $\sigma_r$ of the clouds, such that $Q \equiv \sigma_r \kappa / 3.36 G \mu \gtrsim 1$ (Toomre 1964). Here $\kappa$ is the epicyclic frequency, and $\mu$ is the surface density of the disk. For the case when the disk harbours all the matter necessary to generate a flat rotation curve, both $\kappa$ and $\mu$ scale as $R^{-1}$ and one finds $Q \simeq 0.6 \sigma_r / \sigma_z$ for a flattening of 0.1. Preventing global disk instabilities for flat disks (Toomre 1981) requires larger values of $Q \gtrsim 3$. This suggests that the distribution of dark matter should be significantly rounder than a flattening of 1:10.

Investigations of the stability of rapidly rotating, radially cold oblate spheroidal mass distributions have shown that such models are unstable to a ring instability if as flat or flatter than E6 (4:10), and to an $m = 1$ instability even when nearly round (Merritt & Stiavelli 1990, de Zeeuw & Schwarzschild 1991). However, these instabilities can be suppressed with sufficient radial pressure. Radially anisotropic E6 oblate isothermal models appear to be unstable to bar-like modes (Levison et al. 1990), but models near the isotropic rotator limit with this ellipticity appear to be stable. Instabilities like these have long been suspected to prevent the existence of elliptical galaxies with flattenings greater than E6 or E7.

### 4.2 Distribution of dark matter clouds

We assume that the distribution of dark matter clusters is roughly spheroidal with a flattening of 1:3. Besides the stability arguments, this is also motivated by the following:

(i) Analysis of the temperature and thickness of the thin H I layer in the outer disks of several spiral galaxies including our own shows that most of the dark mass must be located well outside, and with a much greater scale height than, the disk (van der Kruit & Shostak 1984, Merrifield 1992).

(ii) The rotation curves of polar ring galaxies enable the gravitational potential of the dark halo to be sampled along both the minor as well as the major axes of the dominant galaxy. Recent analyses indicate relatively flattened haloes around some S0 galaxies (Rix 1995). E.g, in NGC 4650A (Sackett et al. 1994b) an acceptable range for the halo flattening is between E5-E7. The fact that this is very similar to the observed flattening of the bulge of NGC 4650A is a further interesting result that may support a similar dynamical origin.

(iii) The kinematics of nearby Population II stars are suggestive of a flattened population (White 1985), which, before the disk formed, might have been rounder, perhaps resembling an E3 galaxy (Binney & May 1986). The dark halo, if baryonic, could presumably have formed coevally with this early stellar population, and if baryonic, would then plausibly be more flattened than E3.

(iv) Sancisi (1988) suggested a two-component structure in the HI distribution of isolated galaxies. The HI consists of a thin (thickness $\sim 300$ pc) inner component, which extends radially to about the edge of the optical disk, and a lower density, thick envelope (vertical extension up to $\sim 5$ kpc), which extends radially beyond the optical disk. This second component has a density distribution which falls off more slowly with radius than that of the inner component and may trace the baryonic dark matter halo. At a typical optical radius of $\sim 15$ kpc, this observed component would still be rather flattened; however, this may in part be an effect of the proximity of the active, star-forming disk.

(v) An isotropic or radially anisotropic oblate spheroidal halo damps out the warp modes of self-gravitating disks (Nelson & Tremaine 1995) over a time-scale short compared to a Hubble time unless the halo is very nearly spherical.

This raises the problem of a plausible excitation mechanism operating throughout a typical warped galaxy's lifetime. On the other hand, halos with velocity distributions biased towards near-circular orbits, such as might be expected if the halo dark matter is baryonic and partially dissipative, can excite warps; there would then be no difficulty with the large frequency of observed warps.

We note that if galaxies with E7 flattened halos were to merge, the resulting halos would be less flattened than those of isolated galaxies.

We assume that the mean distribution of clouds defines a mass density $\rho(R,z) = M n(R,z)$, where $n(R,z)$ is a spatial probability density and $M$ is the mean mass of the clouds. The cloud halo is modelled by an oblate spheroid with axis ratio $e$ and a singular isothermal density law. Taking $v_c = 200 v_{200}$ km s$^{-1}$ to be the asymptotic circular velocity in the disk, the cloud ensemble has density

$$\rho(R,z) = \frac{v_c^2}{4\pi G e \alpha}(R^2 + \frac{z^2}{e^2})^{-1}, \qquad (18)$$

and face-on projected density

$$\mu = v_c^2/4\alpha GR = 230 \, \alpha^{-1} R_4^{-1} v_{200}^2 \, M_\odot/\mathrm{pc}^2. \qquad (19)$$

Here $R_4$ is the radius in units of 10 kpc, and the factor $\alpha = 1.33 - 1.2$ for $e = 0.3 - 0.5$. Solving the Jeans equation for an isotropic rotator model with this mass distribution, one finds that the velocity dispersion in the equatorial plane is given, to within less than 10%, by

$$\sigma \simeq 1.16\sqrt{e} \, (v_c/\sqrt{2}) \qquad (20)$$

for flattening $0.05 < e < 0.5$. For a near-spherical halo, $\sigma \lesssim v_c/\sqrt{2}$.

## 5 HALO GAS CLOUDS EMBEDDED IN MACHO CLUSTERS

Machos are presumed to form in dense clusters. Stars form in dense clusters, and there is no reason to expect that Macho formation would differ. The cluster mass range is determined by the initial conditions describing the gas in the protogalactic halo. Naive extension of the Fall-Rees (1985) derivation of globular cluster masses, due to thermal instability at $\sim 10^4$K, suggests that a subsequent thermally unstable phase will be induced at anywhere from $\sim 100$K, if the gas is not enriched above the old Population II abundance level ($Z \sim 10^{-3} Z_\odot$), to the cosmic microwave background temperature at the appropriate redshift, (*e.g.*, $T \sim 15$K at $z \sim 4$). One would certainly expect halo formation to be initiated by the epoch of quasar formation, when quasars, identified with the nuclei of galaxies, and damped Lyman alpha absorption clouds, identified with early stages of disk formation, are in place. The inferred cloud mass in this cold gas phase of halo evolution, assuming pressure equilibrium with the hot halo gas, is $\sim (T/10^{6.5}\mathrm{K})^2 10^{11} \mathrm{M}_\odot$, or $2 - 100 \mathrm{M}_\odot$. The clouds are gravitationally unstable to continuing fragmentation. Their ultimate fate, we argue, incorporates preferential formation of Machos near the fragmentation limit, $\sim 10^{-3} - 10^{-2} \mathrm{M}_\odot$, in which case the primordial clouds fragment into "miniclusters" of Machos and residual gas, which we subsequently refer to simply as Macho clusters.

The Machos themselves develop into non-dissipative clusters, within which residual gas can be stably embedded. Unlike the situation in open clusters or in globular clusters, stellar feedback provides no heat input into the gas, and (*c.f.* the idealized calculations in Section 3) up to $\sim 30$% of the cluster mass can remain gaseous and avoid fragmentation into stars. We now investigate a model in which substantial amounts of cold gas are hidden in dense clouds embedded within Macho clusters, and derive constraints on their physical parameters and distribution.

We characterize a single gas cloud by its mass $M$, effective surface area $A$, and optical depth $\tau$, so that its mass column density is $M/A$. One constraint on the cloud parameters may be obtained by demanding that cloud collisions dissipate energy on a sufficiently long time scale. Specifically, we assume that the collision time $t_c = (nAv_{\mathrm{rms}})^{-1}$ equals a Hubble time near the radius of the optical disk. Within this radius, dissipation will have been significant in assembling the inner disk. Outside this radius, the halo gas has not yet had time to settle into the disk plane. Since for an isothermal sphere model, the density is known from the rotation velocity, this is a direct constraint on the specific cloud cross section $A/M$, ie. on the cloud column density.

Consider the following model for a population of identical cold clouds and Macho clusters that jointly constitute the dark matter in the halo. The cloud column density is

$$N_H = 10^{23} \, \chi_g \, f_{0.01}^{-1} \, \mu_{10} \, \mathrm{cm}^{-2}, \qquad (21)$$

where $\chi_g$ is the mean gas fraction and we normalize to a fiducial dark matter surface density $\mu_{10} \equiv \mu/10 M_\odot \, \mathrm{pc}^{-2}$ and to a disk surface covering factor $f_{0.01} \equiv f/0.01$. We require that the clouds should not collide with one another within the age of the galaxy, taken to be $t_G = 15 \, t_{15}$ Gyr. We further constrain the clouds to be marginally gravitationally bound within the Macho clusters, and the cloud thermal evaporation time-scale to exceed the age of the galaxy. The clouds are required to also be dynamically stable, and the Macho clusters are required to survive dynamical evaporation.

### 5.1 Collisions

The rms cloud velocities perpendicular to the disk are $v_{\mathrm{cl}} \approx 0.8 v_c \, (H/R)^{\frac{1}{2}}$ according to equation (20) for an oblate halo with semi-minor axis $H$ and semi-major axis $R$. The collision constraint implies that

$$f^{-1} H / \sqrt{3} v_{\mathrm{cl}} > t_G, \qquad (22)$$

or

$$f_{0.01} < 0.26 \, R_4 \, v_{200}^{-1} \, e_{0.3}^{\frac{1}{2}} \, t_{15}^{-1} \qquad (23)$$

where $e_{0.3} \equiv (H/R)/0.3$ is the halo flattening, and the factor $\sqrt{3}$ in equation (22) accounts for random collision directions of clouds with isotropic velocities. Using eqs. (19),(21) this becomes a constraint on the gas column density for cloud survival,

$$N_{23} > 90 \chi_g \, R_4^{-2} \, v_{200}^3 \, e_{0.3}^{-\frac{1}{2}} \, t_{15}. \qquad (24)$$

At an optical radius of 15 kpc appropriate to NGC 3198 or our Galaxy, $f < 0.004 v_{200}^{-1}$, the surface density of the dark halo is $\mu = 150 v_{200}^2 \, \mathrm{M}_\odot/\mathrm{pc}^2$, and the constraint on the column density becomes $N_H > 4 \times 10^{24} v_{200}^3$ cm$^{-2}$. Such clouds would also be rather robust against shocking by the interstellar medium in the Galactic disk. Since this has an average column density of $5 \times 10^{20}$ cm$^{-2}$, much less than 1% of their momentum should be lost in an average passage through the disk.

### 5.2 Hydrostatic support

As shown above, the gas clouds can be stabilized by the Macho clusters if they are more centrally concentrated than the Macho distribution. The clouds will be in near-hydrostatic equilibrium, and it is reasonable to assume that they marginally satisfy the Jeans criterion to within a factor of two or so. For a cloud of density $10^7 \, n_7$ cm$^{-3}$ and temperature $10 T_{10}$ K, this condition requires that the cloud radius

$$L = \lambda_J/2 = 0.017 \, g^2 \, T_{10} \, N_{23}^{-1} \text{ pc} \tag{25}$$

and the cloud mass

$$M = 0.3 \, g^3 \, T_{10}^{\frac{3}{2}} \, n_7^{-\frac{1}{2}} \, M_\odot = 0.85 \, g^4 \, T_{10}^2 \, N_{23}^{-1} \, M_\odot, \tag{26}$$

where $g = \Delta V/c_s$ is the ratio of turbulent velocity $\Delta V$ to thermal velocity $c_s$ within a cloud, and $N_{23}$ is the column density in units of $10^{23}$cm$^{-2}$.

### 5.3 Evaporation

In most circumstances, the halo clouds will be embedded in an atmosphere of hot gas, supported by thermal pressure at the virial temperature (Spitzer 1956): $k \, T_h/\mu_H \, m_p = \sigma^2 = \frac{1}{2} v_c^2$ or $T_h = 1.5 \times 10^6 \, v_{200}^2$ K, where $\mu_H = 0.61$ is the mean atomic weight of the ionized H, He gas. We expect $v_c \approx 200$km s$^{-1}$. If the hot gas is in pressure equilibrium with the interstellar medium in the disk, it will have density roughly $10^{-3}$ cm$^{-3}$ and total mass $\sim 10^8$ M$_\odot$ within 10 kpc. The hot gas transfers energy to the surfaces of the cooler clouds by thermal conduction (Cowie & McKee 1977). The sheath of warm gas forming around each cloud will be lost from the cloud and added to the halo gas. Since the clouds also move at nearly the virial temperature, the energy per unit mass in the halo does not decrease appreciably by this process. Thus one may imagine a quasi-steady state in which the halo loses mass to the disk on a cooling time scale ($\sim 5 \times 10^9$ yr), gains mass from the dense halo clouds through evaporation, and in addition gains mass and energy from the disk through multiple supernova–driven galactic fountains. However in the outer halo, where the bulk of the dark mass resides, there is no energy or mass input to the halo gas from the disk. The clouds will evaporate, and the residual amount of hot diffuse gas will be determined by the gas cooling timescale. This exceeds the disk age only if the diffuse halo gas has density $n_h \lesssim 10^{-4}$ cm$^{-3}$. The clouds may also be heated by the soft X-ray background.

The evaporation rate of a spherical cloud in a medium of temperature $10^6 \, T_6$ K is saturated if the intercloud medium density satisfies $n_h < 4 \times 10^{-3} \, L_1^{-1} \, T_6^2$ cm$^{-3}$, a condition that is easily satisfied for $T_h > 10^6$ K. Here we denote the cloud radius by $L = 1 L_1$ pc. The resulting evaporation rate, in the limit of saturated thermal conductivity (Cowie & McKee 1977), from a spherical cloud is

$$\dot M = 5 \times 10^{-4} n_h T_6^{\frac{1}{2}} L_1^2 \text{M}_\odot \text{ yr}^{-1} \tag{27}$$

for a saturation parameter of 40, valid for $n_h \sim 10^{-4}$ cm$^{-3}$. For fixed column density, this decreases strongly with the radius of the cloud. Requiring that the clouds not evaporate within the age of the galaxy imposes the constraint

$$N_{23} > 0.2 n_{-4} T_6^{\frac{1}{2}}. \tag{28}$$

The collision constraint on the cloud column density therefore guarantees that the clouds survive evaporation. The total mass loss rate by evaporation is

$$\dot M_{\rm tot} = 10 n_{-4} T_6^{\frac{1}{2}} N_{23}^{-1} \chi_g M_{12} \text{ M}_\odot \text{ yr}^{-1}, \tag{29}$$

where the total halo mass is $M_{\rm tot} = 10^{12} M_{12}$ M$_\odot$. Coincidentally, this is comparable to the present day disk star formation rate, thereby providing a possible means of gas replenishment.

### 5.4 Thermal balance

The dominant heat source for gas clouds at large galactocentric radii may be cosmic rays or the diffuse X-ray background. Halo soft X-rays do not penetrate sufficiently deep enough into clouds with column density in excess of the required $10^{23}$ cm$^{-2}$.

The spectrum of the X-ray background in the energy range $2.5 < E < 25$ keV is

$$I(E) = 8.5 E^{-0.4} \text{keV cm}^{-2}\text{s}^{-1}\text{keV}^{-1}\text{sr}^{-1}, \tag{30}$$

where $E$ is measured in keV (Longair 1995). The integrated cross section for absorption of X-rays by gas with cosmic abundance composition is

$$\sigma_X = 2.6 \times 10^{-22} E^{-8/3} \text{ cm}^2 \tag{31}$$

(Morrison & McCammon 1983, O'Dea *et al.* 1994); thus the typical energy of an X-ray photon heating the core of a cloud with column density $N = 10^{23} - 10^{24}$ cm$^{-2}$ is $3 - 8$ keV. Higher-energy photons penetrate the cloud essentially unimpeded, while the low-energy photons only heat the cloud surface. Integrating the cross section over the X-ray background spectrum, the ionization rate per H$_2$-molecule from secondary ionizations is (Voit 1991)

$$\xi \simeq 26 \int_3^{10\text{keV}} 4\pi I(E)\sigma_X(E)\, dE = 4 \times 10^{-20}\text{s}^{-1}. \tag{32}$$

If half of the corresponding energy input goes into heating, the heating rate per H$_2$ molecule is

$$\frac{\Gamma}{n(\text{H}_2)} = 0.5\xi \times 15.4\text{eV} = 5 \times 10^{-31} \frac{\text{ergs}}{\text{s} \cdot \text{H}_2}. \tag{33}$$

This is a very low heating rate compared to that by cosmic rays in the galactic disk.

Clouds with primordial composition cannot cool below $\gtrsim 100 K$ (Hollenbach & McKee 1979, Lepp & Shull 1984, Murray & Lin 1990). With a column density of $N_H = 10^{24}$ cm$^{-2}$, such clouds in the outer halo must then have masses $\gtrsim 10$ M$_\odot$, densities $\sim 10^7$ cm$^{-3}$, and radii $\sim 0.02$ pc. However, it is likely that during the early phases of galaxy formation even gas in the outer halo was contaminated to small but non-zero metallicity; for compositions characteristic of extreme Population II, $[Fe/H] > -3$, cooling by metals will lead to more plausible temperatures, of $\sim 5 - 20$ K (de Forets *et al.* 1995).

To determine the equilibrium cloud temperature, we require the cooling rates for molecular gas clouds. Calculations which incorporate the opacity of the gas in a simple velocity gradient radiative transfer model (Goldsmith & Langer (1978; Neufeld *et al.* 1995) find that at the high densities under discussion here, $n_7 \sim 1$, the cooling rate per molecule actually *decreases* with density. Note that in absolute value, however, at these very high densities, the newer calculations find a cooling rate about 100 times as large as in the older study, due to revised molecular excitation rates. From the diagrams in Neufeld *et al.* we estimate

$$\frac{\Lambda}{n(\mathrm{H}_2)} = 2.5 \times 10^{-27} n_7^{-0.3} T_{10}^{2.6} \frac{\mathrm{ergs}}{\mathrm{s} \cdot \mathrm{H}_2}. \tag{34}$$

This is valid for a cosmic ray ionization rate of $\xi_{\mathrm{CR}} = 10^{-17}\,\mathrm{s}^{-1}$, typical of the inner Galaxy, and for standard galactic disk gas phase element abundances. Unfortunately, calculations like those of Neufeld *et al.* are not available for the lower element abundances and cosmic ray ionization rates anticipated in the outer halo. Thus the expected cloud equilibrium temperature is uncertain on account of both $\Lambda$ and $\Gamma$.

For illustration, if we assume that cloud heating occurs from ionization at a rate of $\xi = \xi_{18} 10^{-18}\,\mathrm{s}^{-1}$, intermediate between that expected from the X-ray background and from cosmic rays in the inner Galaxy, and balance this by the galactic disk cooling rate (34), we obtain a formal equilibrium temperature of

$$T_{10} \simeq 0.12 n_7^{1/10} \xi_{18}^{0.4} \Lambda_N^{0.4}, \tag{35}$$

where $\Lambda_N = 1$. In other words, the clouds would in this case have cooled down to the microwave background temperature. The clouds could be warmer if either the actual cooling rate was lower than the Neufeld *et al.* value on account of lower molecule abundances and lower ionization rates, or if the heating rate were larger.

Eq. (35) has another interesting consequence: if $T_{10} \sim 1$, we expect a polytropic equation of state with $n \simeq 10$:

$$p \propto \rho^{\frac{11}{10}}. \tag{36}$$

(For the Goldsmith & Langer scaling of $\Lambda$ we would have found $n=7$: *c.f.* Section 3).

### 5.5  Stability of the Macho cluster

The collision constraint on gas column density, combined with the assumption of near-Jeans hydrostatic equilibrium, implies a cloud density

$$\frac{3M}{4\pi L^3} > 4.1 \times 10^4 \,\mathrm{M}_\odot\,\mathrm{pc}^{-3} N_{23}^2 T_{10}^{-1} g^{-2}. \tag{37}$$

The stability considerations of Section 3 suggest that the mass of the Macho cluster must be greater than about three to five times the mass of the embedded gas cloud, and its radius greater than about three times the mean radius of the cloud. From eqs. (26)), (25)) this results in

$$M_d \gtrsim 4 g^4 T_{10}^2 N_{23}^{-1} \,\mathrm{M}_\odot, \tag{38}$$

$$r_d \gtrsim 0.05 g^2 T_{10} N_{23}^{-1} \,\mathrm{pc}. \tag{39}$$

If we assume that the individual Machos have masses near $\sim 10^{-2}\,\mathrm{M}_\odot$ as one might expect from opacity-limited fragmentation (Silk 1978), the cluster half-mass relaxation time is

$$t_{rh} \gtrsim 1 \times 10^8 M_{d10}^{1/2} r_{d1}^{3/2} (m_*/10^{-2}\,\mathrm{M}_\odot)^{-1}\,\mathrm{yr}, \tag{40}$$

where $M_{d10} = M_d/10\,\mathrm{M}_\odot$, $r_{d1} = r_d/1\,\mathrm{pc}$. Dynamical evaporation takes about 300 half-mass relaxation times (Spitzer & Thuan 1972). Thus the requirement that the Macho cluster is stable against evaporation for $10^{10}$ yr implies

$$t_{\mathrm{ev}} \gtrsim 2 \times 10^8 g^5 T_{10}^{5/2} N_{23}^{-2} (m_*/10^{-2}\,\mathrm{M}_\odot)^{-1}\,\mathrm{yr} > 10^{10}\,\mathrm{yr}, \tag{41}$$

so that the ratio $g$ of macroscopic to thermal velocity must satisfy $g \gtrsim 2$ for the specified parameters. At a given Macho mass $m_*$, this condition becomes weaker for larger Macho clusters around a given gas cloud.

One can live with Machos of mass $\sim 10^{-2}\,\mathrm{M}_\odot$ if $T_{10} \sim 2$ and $g \sim 1.5$, or if $T_{10} \sim 1$ and $g \sim 2$. Machos of mass $\sim 0.1\,\mathrm{M}_\odot$ would be excluded, but a mass range so close to the M dwarf limit seems very contrived, if not already excluded on observational grounds from the direct searches that constrain the abundance of extreme halo M dwarfs. An M dwarf spillover to the mass range just above $\sim 0.1\,\mathrm{M}_\odot$ would seem inevitable. Most of the required macroscopic motions may be in rotation, which is likely to be more stable to dissipative evolution than pure turbulence. If the clouds were rotating, then their relative mass fraction could be somewhat greater than we have estimated in Section 3 from the spherical model.

### 5.6  Evolution with galactocentric distance

Consider now the evolution of the gas content of the halo with distance from the center of the galaxy. If we choose a fiducial value of $N_{23} = 1$ for the column density of the gas clouds, the ratio of cloud mass to the mass of the surrounding Macho cluster could be somewhat less than unity, and, from the previous discussion one may argue that this ensures both stability of the cloud and survival of the Macho cluster. From eq. (24) we can solve for the fraction of gas remaining in the halo clouds as a function of time and radius:

$$\chi_g = 0.01 N_{23}\, R_4^2\, v_{200}^{-3}\, e_{0.3}^{\frac{1}{2}} t_{15}^{-1}. \tag{42}$$

At 100 kpc, the halo can thus contain a substantial fraction of cold gas, yet near the Sun will by now have lost most of its gas, with only the clustered Machos remaining. Suppose that cloud collisions indeed liberate gas up to the maximum amount allowed by the cloud orbits. In this case, the gas mass in the inner galaxy decreases inversely with galactic age, so that the gas accretion rate onto the disk arising from halo cloud destruction decreases with time, according to $t^{-2}$. Inclusion of gas accretion is of considerable importance for models of galactic chemical evolution, and is a natural consequence of our dark matter model.

Finally, we note from eq. (42) that, at fixed radius and cosmic epoch, the halo gas content scales as $v_{200}^{-3}$. For example, a dwarf galaxy with $v_c = 70\,\mathrm{km\,s}^{-1}$ should have 25 times higher remaining halo gas fraction in its outer halo than a normal spiral galaxy like ours. Correspondingly, its luminosity-to-dark mass ratio should be distinctly lower. If the assembly of the disks of spiral galaxies were driven mainly by accretion from their baryonic halos, this would naturally lead to a picture in which giant galaxies evolve on the fast track, while small galaxies still appear to be underluminous and dark matter–dominated.

# 6 OBSERVABLE PROPERTIES OF TWO-COMPONENT CLOUDS

Other observable properties of such clouds at the peripheries of galaxies include patchy extinction on sub-arcsec scales towards globular clusters and distant galaxies, absorption both of clouds in our halo and of clouds in intervening halos against distant quasars, OVI absorption, 21cm absorption and emission, molecular (CO, $H_2O$) absorption or emission, millimeter-wave emission from cold dust, and X-ray absorption. With the low surface covering factor of $\lesssim 1\%$ for our canonical cloud parameters, none of these probes is as yet particularly restrictive.

## 6.1 Gravitational microlensing in the Galactic halo

The handful of gravitational microlensing events detected to date towards the Large Magellanic Cloud requires Machos in the mass range $0.01 - 0.1\,M_\odot$. Halo flattening, as motivated by a baryonic halo model, tends to increase the mass fraction in Machos for a specified number of microlensing events and rotation curve (Alcock *et al.* 1995b). The inferred halo mass fraction in a flattened halo is about 50%, but with large uncertainty. The somewhat lower mass Machos preferred by our model should be detectable on microlensing timescales of order 10 days, well within the range of ongoing searches.

## 6.2 Obscuration of background stars in microlensing experiments

A gas cloud with typical size given by eq. (25) and typical transverse velocity $v_c$ traverses its own radius in $\sim 80 g^2 T_{10} N_{23}^{-1} v_{200}^{-1}$ yr. A background star in the Magellanic clouds that happens to be obscured behind one of the halo clouds would reemerge after about this time interval. For a cloud area filling factor of $\lesssim 0.01$ (see eq. (23)), we expect $\lesssim 1\%$ of background stars to be presently obscured. The angular extent of typical halo clouds at distance 10 kpc is about a third of an arcsec; thus background stars would be individually obscured. On average, the number of background stars that become newly obscured per unit time is equal to the number of stars that reemerge. For a total of a million Magellanic cloud stars monitored by the Macho experiments we therefore expect $\lesssim 10^4$ stars to be obscured and reemerge at a rate $\lesssim 100/$ yr. Search for such events in the Macho samples should therefore provide a stringent test of our model.

## 6.3 Young halo stars and pulsars

The relatively rapid dynamical evolution of low mass Macho clusters means that those rare objects which only marginally satisfy the survival condition eq. (41) would merge and possibly form stars in the halo. This could provide a means of occasionally forming apparently isolated main sequence stars far from the galactic disk. The deaths of these stars would result in halo supernovae and formation of pulsars with Population II kinematics.

Preston, Beers & Shectman (1994) found a population of metal-poor ([Fe/H]$\lesssim -1$) A- and F-type stars with kinematic properties unlike any known galactic population. These stars rotate around the Galaxy with $v_\phi = 128\,\mathrm{km\,s^{-1}}$ and have an isotropic velocity distribution with $\sigma \simeq 90\,\mathrm{km\,s^{-1}}$, intermediate between values for the thick disk and metal-poor halo. Their space density within 2 kpc of the Sun amounts to $4 - 10\%$ of the density of the metal-poor halo. Preston *et al.*'s interpretation of this population, containing $\simeq 10^8\,M_\odot$, is in terms of multiple accretion of star-forming satellite galaxies within the past $\gtrsim 3$ Gyr. An obvious alternative in the context of our model is that these stars have formed over the past few Gyr from clouds *within* the Milky Way halo. In this context it is interesting to note that equation (20) predicts a velocity dispersion of $100\,\mathrm{km\,s^{-1}}$ in the galactic disk plane for an isotropic oblate halo model of ellipticity $e = 0.3$ and a circular rotation velocity of $v_c = 220\,\mathrm{km\,s^{-1}}$, very close to the observed values.

There are also B-type stars observed in the galactic halo at $z$–distances of several kpc, some of which *cannot* have formed in the galactic disk and subsequently been ejected (Conlon *et al.* 1992). These stars have approximately solar metallicity and must thus originate from more enriched gas. It would be interesting to investigate the kinematics of these stars in more detail, since they would not have had time to phase-mix after a possible accretion event. Proper motion studies could eventually discriminate between an accretion origin and the hypothesis that these stars could have formed from halo gas clouds.

## 6.4 HI / CO emission and absorption

We have implicitly assumed above that most of the material in the halo clouds would be molecular hydrogen. In primordial clouds, $H_2$ can be formed by $H^-$ and $H_2^+$ processes, but the $H_2$ fraction remains at $\lesssim 10\%$ (Lepp & Shull 1984). Since the optical depth in HI at a column of $10^{24}\,\mathrm{cm^{-2}}$ and a cloud temperature of $10 - 100$ K is $\sim 10^6 - 10^5$, their smeared-out brightness temperature at 21cm would be $0.1 - 1$ K for $f \sim 0.01$. This results in a weak emission signal broadened to a velocity dispersion of order $100\,\mathrm{km\,s^{-1}}$ (eq. (20)). A similar signal is expected from the HI envelopes resulting from heating of the cloud envelopes by halo soft X-rays.

If, on the other hand, the clouds contain dust, they will be molecular at this column density and not emit in HI. However, there would then be substantial amounts of CO. This would be highly beam–diluted in any single dish emission experiment. Typical clouds at $\sim 10$ kpc distance are likely to subtend an angle of $\lesssim 0.4$ arcsec. With $\sim 200$ clouds in a beam of 1 arcmin, one might expect $\sim 10$ percent beam–to–beam fluctuations in the CO emission with a mean brightness temperature of $\sim 0.1$ K, again provided that one can resolve the broad wings of $\sim 100\,\mathrm{km\,s^{-1}}$ width. This is an upper limit because the inner halo is likely to be depleted of gas.

Lequeux, Allen and Guilloteau (1993) have recently reported the detection of cold CO in absorption in the Perseus arm, at a galactocentric distance of 12 kpc. These authors note that in at least one case there is no corresponding detection of HI, and infer that the ratio of $H_2$ to HI column densities at this distance may be about 4 or so, as compared to a value of roughly unity near the sun.

During galaxy formation, a process that, like extreme starbursts, is plausibly associated with mergers between massive clouds, the gas may have been detectable via $CO$ and $H_2O$ emission. There have been recent reports of pre-liminary detection of molecular emission from damped Lyman alpha clouds at high redshift (Brown & Vandenbout 1993, Encrenaz *et al.* 1993) at a level that, if confirmed, is suggestive of a large reservoir of molecular gas. Damped Lyman alpha clouds are often identified with the high redshift precursors of disk galaxies in the early phase of a forming galaxy. Frayer, Brown and Vanden Bout (1994) detect CO (1 to 0) and CO (3 to 2) emission and infer approximately $10^{12} M_\odot$ of molecular gas in one such system at $z = 3.4$. The higher CMB temperature and enhanced XRB heating means that the gas clouds in our model are likely to be prolific emitters in CO at $z \sim 3$.

### 6.5  Diffuse gamma ray emision

Cosmic rays penetrate the cold molecular globules and produce gamma rays via nuclear interactions with hydrogen and helium nuclei. The diffuse gamma ray emission from the inner galaxy provides an excellent probe of cold molecular hydrogen (Strong *et al.* 1988), measuring $\sim 1 \times 10^9 M_\odot$. Observations at high galactic latitude are less constraining (Bloemen 1989). However, the following crude estimate suggests that gamma rays are a potentially important probe of our hypothesised clouds: with a filling factor of 1 percent, 100 times the mass in known interstellar gas, about $4 \times 10^9 M_\odot$, would give an equal contribution in gamma rays to that observed. Indeed, a detailed study of cosmic ray diffusion models in the halo concludes (Salati *et al.* 1995) that no more than about 9% of halo dark matter can be in the form of gas, whether in diffuse form or in clouds. However as the inner halo is depleted of gas, the gamma ray limit weakens: for example with a deficiency of gas within 2 kpc above the disk, the limit on halo gas increases to about 50%. It is apparent that one could hide very substantial amounts of cold gas, especially in the outer halo where the cosmic ray flux is likely to be lower than in the inner galaxy. High angular resolution gamma ray observations, to search for a clumpy diffuse component at high galactic latitudes, are needed to test our model.

### 6.6  Millimeter wave continuum emission from cold dust

Cold molecular gas clouds may be indirectly observable via dust emission. At the periphery of the galaxy, where there is little starlight, the dust most likely will be cold. One might expect the dust temperature at radius $r$ to be of order $20(r_{\rm opt}/r)^{2/(4+\beta)} (0.1\mu/a)^{\beta/(4+\beta)}$ K, where $r_{\rm opt}$ denotes the half-light radius, $a$ is the typical grain radius, and $\beta \approx 1-1.5$ is the dust emissivity index. The dust should therefore be at a few (3-5) degrees Kelvin, at, say, 50 kpc from the center of the galaxy, especially if the dust particles that form in the quiescent cold clouds by coagulation over a Hubble time are larger than nearby dust in star-forming clouds (Wright 1993). This might have dramatic implications for experiments that are searching for cosmic microwave background fluctuations in the millimeter and submillimeter spectral regions. Foreground emission from galactic dust is an important contaminant, that can be modelled out by higher frequency IRAS observations in terms of a galactic dust component at $T \sim 20$ K. Cooler, patchy dust emission at high galactic latitudes, where some of our clouds will inevitably reside, would present a source of noise that might be very difficult to remove from the cosmological background signal. Indeed, at least one recent experiment (Cheng *et al.* 1993) that finds fluctuations attributed to the cosmic background may also have detected unresolved (at 20 arc-min resolution) "sources" whose spectral characteristics are indistinguishable from that of a blackbody at $\lesssim 4$ K. Of course these features may also be explained by correlations intrinsic to the cosmic microwave background (Kogut *et al.* 1995), but the principle remains: one should carefully search for high galactic latitude cold dust-like signals. Indeed, Puget *et al.* (1995) have reported the detection in the FIRAS data of a far infrared background signal, consistent with an early universe origin, but also conceivably with a halo model such as the one we are advocating here. One could test such a hypothesis by examining nearby galaxy halos for evidence of diffuse far infrared emission.

### 6.7  Irradiation by UV background and $H_\alpha$ emission

The clouds are exposed to a diffuse ionizing radiation background below 912A which amounts to (Madau 1992) $\sim 6 \times 10^{-24}$ erg cm$^{-2}$ s$^{-1}$ Hz$^{-1}$ sr$^{-1}$. This is appropriate at low redshift; at $z \approx 2-3$, the ionizing flux is inferred to be a factor $\sim 40$ larger. The total flux of ionizing photons incident on a cloud can be written as $10^4 \phi_4$ cm$^{-2}$ s$^{-1}$ with $\phi_4 \sim 1$, and produces a sheath of ionized gas of radius $\Delta L = \phi (\alpha_B n^2)^{-1}$ around the neutral cloud, where $\alpha_B$ is the hydrogen recombination coefficient to excited levels. The column of HII is $N_{II} = 6 \times 10^{17} (\phi_4 L_1)^{1/2}$ cm$^{-2}$, the density at the ionization front (Dyson 1988) is $n_i = 0.3(\phi_4/L_1)^{1/2}$ cm$^{-3}$, and the $H\alpha$ emission measure is $\int n_i^2 dr \simeq 0.2 n_i^2 L \sim 0.02 \phi_4$ cm$^{-6}$ pc. This is not easily detectable unless the ionizing flux is locally enhanced by a large factor.

Such an enhancement can indeed occur when one of the dark clouds passes through a Galactic HII region or old supernova remnant. Such a traversal takes about $10^5$ yr, during which the cloud is suddenly exposed to a high photon flux $\phi = 10^{10} \phi_{10}$ cm$^{-2}$ s$^{-1}$, with $\phi_{10} \sim 1$ about 2 pc from an O-star. A D-type ionization front is driven into the cloud. The front propagates subsonically into the cloud at speed (Dyson 1968, Spitzer 1968) $v_0 = \phi/n = 10^3 \phi_{10} n_7^{-1}$ cm s$^{-1}$. While the cloud traverses the HII region, the front moves $\lesssim 10^{-3}$ pc into the cloud. The ionized gas streams away from the cloud at the sound speed of the HII region and produces an ionized sheath of density $n_i \sim 300 \phi_{10}^{1/2} L_1^{-1/2}$ cm$^{-3}$ with associated emission measure $\sim 2 \cdot 10^4 \phi_{10}$ cm$^{-6}$ pc. In a typical HII region, one might expect to find several such bright clouds in projection. Because of their large transverse motions, the ionized sheaths would appear as nearly straight filaments, extending over a fraction of the HII region because the density and emission measure decrease rapidly with time.

However, the Macho clusters are likely to be gas–free in the inner halo, and any gas cloud interaction with HII regions would occur, if at all, in the outer disk at a distance (*e.g.*, in the Perseus arm) perhaps 4 kpc from the sun. In this case, the filaments would extend over only $\lesssim 1$ arcsec width for cold halo gas clouds that are traversing a large HII region. It would be of interest to carefully search for such objects at high resolution.

### 6.8 Absorption

One possible way of detecting dense halo clouds in galaxies is through the absorption of X-rays from a background source such as a galaxy cluster. At a column density of $10^{24}\,\mathrm{cm}^{-2}$, even a pure hydrogen-helium gas is essentially opaque to X-rays in the energy range measured by ROSAT (0.1-1keV) (Gorenstein 1974). Thus the fraction of X-rays absorbed over the area covered by a foreground galaxy halo is directly given, up to a geometrical factor of order unity, by the area filling factor in eq. (22), $0.001 < f < 0.01$. This suggests that thousands of X-ray photons from the background source will be necessary to detect any absorption.

With the Hubble Space Telescope, one can search for absorption lines from clouds in our own halo against the brightest quasars. Too few lines of sight have been studied to give any interesting constraint. Distant quasars can also be studied to search for absorption in the halos of intervening galaxies. However the halo gas may not be predominantly molecular at high redshift. Indeed, substantial amounts of atomic gas are already observed at $z \sim 3$, comparable to that seen in stars at the present epoch (Wolfe 1993).

If the gas clouds contain dust, they would produce absorption at optical wavelengths against high surface brightness extragalactic objects, such as elliptical galaxies. The typical size is $\sim 0.1$ arcsec for clouds at 30 kpc distance in our halo.

### 6.9 Galaxy clusters

Intracluster gas is detected via its thermal x-ray emission. Conservative estimates of the baryon fraction, predominantly hot gas, in rich clusters lead to estimates of between 5 and 20 percent within an Abell radius, $1.5\,h^{-1}$ Mpc (Briel *et al.* 1992, Buote & Canizares 1995). Within 2 Abell radii, there are indications that the baryon fraction rises to more than 50 percent (Mushotzky 1993). Evidently, this suggests that *some* of the dark matter sampled on cluster scales may be baryonic. Cold gas in field galaxy halos plausibly would be heated and evaporate to become hot intracluster gas. The infall of a galaxy into the intracluster medium (ICM), which is a factor $\sim 100$ hotter and a factor of $\sim 10$ denser than the hot galactic interstellar medium, would accelerate the evaporation rate by a factor $\propto n\,T^{\frac{1}{2}} \simeq 100$ (*c.f.* equation (28)). In this case, a large fraction of the outer halo gas would be added to the ICM in a few cluster dynamical times, and become visible via X-ray emission.

Ellipticals should be surrounded by similar gas-rich dark Macho clusters, since mergers do not destroy outer clusters by changing orbits. But in galaxy clusters, the enhanced temperature and density of the intracluster gas relative to the halo gas considered previously helps to evaporate the gas. Much of this outer halo gas would remain at large cluster radii where an orbiting galaxy spends most of its time. It may be possible in this way to account for the fact that the mass fraction of the X-ray emitting gas increases with cluster radius. This scenario also predicts that the intracluster gas should be significantly less enriched near the cluster boundary than further in. The observed enrichment to about one-third of solar abundance may support the contention that at least some of the cold gas clouds we invoke for the inner galaxy halos are also enriched. Gas around ellipticals and in groups of galaxies reveals lower enrichment than found for intracluster gas, perhaps indicative of a more primordial origin, as expected if the cold gas clouds have survived in less extreme environments.

The x-ray spectra of several galaxy clusters show evidence for self-absorption by relatively cold gas (White *et al.* 1991). The total mass of absorbing gas amounts to $\sim 3 \times 10^{11}\,\mathrm{M}_\odot$, assuming that it is enriched in oxygen relative to iron by a factor $\sim 3$, and is empirically constrained to be in the form of small, cold clouds. In the case of Abell 478, the excess absorption is equivalent to a column of $\sim 3 \times 10^{21}\,\mathrm{cm}^{-2}$ over the central $\sim 300$ kpc of the cluster, which coincides with a region where spectral fitting shows that there is significantly cooler material than the ambient intracluster gas (Allen *et al.* 1993; but see O'Dea *et al.* 1994).

## 7 CONCLUSIONS

Dark halos may well be baryonic, and much of the dark mass within them may plausibly be in the form of clusters of substellar objects within which is embedded cold gas. Typical cluster masses are $\sim 10\,\mathrm{M}_\odot$, typical Macho masses are $\sim 0.01\,\mathrm{M}_\odot$, and the gas content may be up to of order 50 %. Several lines of indirect evidence point in this direction.

We have shown that, once a sufficient fraction of a gas cloud has been converted into such Jupiter-like Machos, the residual gas can be stabilized even for physical parameters that would render it unstable in isolation. If in view of the results of the Macho experiments the fraction of gas remaining is assumed significant, then simple considerations restrict the parameters of the gas clouds to a relatively small region where their temperature is of order $10\,K$ and their column densities are of order $10^{23}\,\mathrm{cm}^{-2}$. For much smaller column densities the clouds will collide, for significantly larger column densities the surrounding Macho clusters will evaporate, and for higher temperatures the gas will become visible.

However, we believe that there are substantial theoretical uncertainties regarding the shapes and dynamical states of these Macho cloud clusters, the expected spectrum of Macho masses, and how variation of these properties affects the dynamical evolution. These questions undoubtedly merit further work; but at the same time, it will be worthwhile to look for and further limit the parameters of these objects observationally.

We have shown that while observations at present provide only a poor discriminant of our hypothesis, a modest improvement in search techniques may well find traces of the clouds that we have postulated to exist. In particular, more sensitive FIR, gamma ray, local and high-redshift mm-line observations, and further analysis of the microlensing experiments will provide tighter constraints on their parameters, and on the baryonic content of dark halos in general.

If indeed a significant fraction of the mass of galactic halos is or once was in the form of cold gas, then our argument about the radial dependence of cloud survival rate has cosmogonical implications for galaxy formation. The halos of galaxies would play a much more active role in what determines the build-up of galactic disks and their chemical evolution, processes which would then naturally occur fastest in massive galaxies.

Merging of substructure, a common aspect of all hierarchical galaxy formation models, would be accompanied by enhanced cloud collision rates and would thus be expected to destroy the reservoir of *inner* halo gas in cold clouds. Outer halos, however, should remain abundant in clusters of Machos and embedded dense cold gas clouds.

## ACKNOWLEDGEMENTS

We are grateful to L. Blitz, K. Freeman, M. Rees, R. Sancisi and R. Wyse for critical and helpful discussions and for comments on an early version of this paper. O.E.G. was supported by a Heisenberg fellowship at Heidelberg, and by a grant from the Schweizerischer Nationalfonds subsequently. The research of J.S. at Berkeley has been supported in part by grants from N.A.S.A. and N.S.F.